# Once again about Einstein's realism: the matter waves and Bell's inequalities from the point of view of the special theory of relativity

Iurii Kudriavtsev

(juku@bk.ru)

*We explore the possibility to give a classical explanation to the specifics and physical sense of de Broglie matter waves when studying the microparticle as an object of non zero size, from the point of view of the special theory of relativity.*

*We show that the particularities of de Broglie matter waves and the results of the experimental verifications of Bell inequalities for the pairs of entangled photons are naturally interpreted as the results of implementation of the conclusions of the special theory of relativity to the microparticles.*

*We conclude that it is appropriate to go back to the search of the new means of realistic description of the nature proposed by Einstein and his realistic worldview that states that the world studied by the science is real and every part of it at any moment of time has objective physical characteristics.*

**03.75.-b**

*«Einstein's general relativity stands out, in my opinion, as that century's greatest single achievement. Quantum theory (and QFT) might well be regarded by most physicists as an even greater achievement. From my own particular perspective on the matter, I do not feel able to share that view»* (Roger Penrose. The Road to Reality)

## 1. Introduction

As is known, Albert Einstein who created the most perfect to the date physical theory and made a great contribution to the microphysics development, had highly appreciated quantum mechanics and the perfection of its mathematical theory but considered its physical interpretation quite unsatisfactory because it contradicted the foundation of his physical worldview that can be described by one short phrase – the world is real:

*«Concepts of physics are related to the real outer world, i.e. they suppose the idea of things that require 'real existence', independent from the perceiving subjects (bodies, fields, etc.); these ideas, on the other hand, are being matched as close as possible with the sensory perception »* [1] (here and further the quotes are translated from the Russian editions).

*«The only acceptable interpretation of Schrodinger equality so far is the statistical interpretation given by Born. However, it does not describe the real state of a separate system and only allows making statistical statements about ensembles of systems.*

*I think that it is wrong to make theoretical ideas the foundation of physics because it is impossible to refuse the opportunity to describe objectively a separate microsystem (i.e. description of the 'real state') without making the physical worldview fade to a certain degree. At the end, it seems unavoidable that the physics must aspire to describe the real state of a separate system. The nature in general can be seen only as a separate (existing on a single occasion) system and not as an 'ensemble of systems'* [2].

*«I do not doubt that the contemporary quantum theory (or more precisely, quantum mechanics) gives the fullest coincidence with the experience, since the foundation of the description as key concepts are material point and potential energy. However, what I find unsatisfactory in this theory is the interpretation that is given to «Ψ-function». Anyway, the basis of my understanding is the concept, strongly rejected by the biggest contemporary theorists:*

*There is something like 'real state' of the physical system that exists objectively, regardless of any observation or measurement that can be described with the help of means that physics possess. [Which adequate means should be used for this purpose, and, respectively, with fundamental notions should be used, is not clear, I think. (Material point? Field? Any other mean that we are still to find?)]»* [3].

The founder of wave mechanics Louis de Broglie was too, looking for an exit from this situation all his life: *«Having started in 1928 my teaching career, I stated some ideas that were prevailing in the quantum mechanics, and was refusing for a long time to develop my own initial ideas. But in about 20 years I understood that it is necessary to go back to the concept of a particle as a very small localized object moving by the trajectory ... I think, my initial ideas which I have gone back to and further developed, give an opportunity to understand a true nature of co-existence of the waves and particles, unlike the usual quantum mechanics and their generalizations that explain it only statistically, without revealing its true content»* [4].

The major concerns of these great physicists of the 20th century are well illustrated by this endlessly sad conclusion of Arthur Haas [5]: *«Looking back at the history of the theoretical physics we see that the essence of the physical progress is in a gradual liberation of physics from the purely human point of views. In this sense the years when the works of de Broglie, Schrodinger, Heisenberg and Dirac appeared, should be considered the period of the clarity that gave to the physics a lot of means to overstep the usual stereotypes».*

One of the first reasons why microphysics were separated from classical physics, realism and *'purely human point of view'* was the unusual character of the matter waves discovered by Louis de Broglie: the length of the wave, inversely proportional to the velocity of the particle, phase speed which exceeds the light speed by as much as the light speed exceeds the velocity of the spatial movement.

Half a century after the firm views on incompatibility of quantum mechanics and classical physics have gotten a new strong confirmation on the results of the experimental verification of Bell inequalities, interpreted as a confirmation of the inconsistency of 'Einstein's realism' [6] and a final failure of the realism conception at all.

We believe that this conclusion from the situation is profoundly wrong, and in the argument of Einstein and the followers of the orthodox interpretation of the quantum mechanics, Einstein is right, when insisting on existence of 'the real state' of the physical system, that is possible to be described, and showing the way to follow – the one of searching the means to describe it [3].

In this work we try to explain the main specifics of the de Broglie matter waves and the results of the experimental verification of Bell inequalities from the point of view of the special theory of relativity to illustrate the possibility to eliminate the gap between the quantum mechanics and 'purely human' point of view, and going back to the realistic worldview.

**2. Length of de Broglie wave and microparticle as an object of non zero size**

Let us see if the idea of impossibility of classical interpretation of de Broglie material waves is related to the common in the first half of 20th century view of microparticles as point objects ('material points'). We can assume that de Broglie himself was insistently going back to the idea of microparticles as very small, localized ('point') objects because he did not presumed that the properties of the space occupied by the microobject can be different from the properties of the space in general. Although such a possibility was mentioned in the works of a variety of physicists, mathematicians, philosophers of $19^{th}$-$20^{th}$ centuries, from Bernhard Riemann to E.J. Zimmerman [7].

Let us imagine the microparticle as an object distributed by a certain area of the space. The question of the nature of this distribution and the size of occupied space is left open. According to de Broglie model, we assume that the microparticle with the mass $m_0$ in own reference system corresponds to the oscillation process with frequency

$$\nu_0 = m_0 c^2/h; \qquad (1)$$

Let us suggest that these fluctuations in the particle's reference system happen synchronously and in phase in the whole volume of the particle as a whole entity with a mass $m_0$. It seems obvious that the assumption of the synchrony and in-phase mode is equivalent to the assumption about the special properties of the microparticle's space.

According to the equalities of the special theory of relativity [8], if the object moves in relation to the immobile observer with the velocity v, the time t in observer's reference system is related to the time t´ in object's reference system by the expression

$$t = \gamma(t´ + v x_0/c^2); \qquad (2)$$

where $x_0$ – the coordinate in own reference system of the object in the direction of the velocity of the movement, c – light speed,

$$\gamma = (1 - v^2/c^2)^{-1/2}. \qquad (3)$$

Let us see the result of the mal-synchronization of the time in the volume of the particle, defined by the second summand in the right part (2). When the point in question declines from the conventional center of the microparticle by the value of $x_0$, from the point of view of observer it results with the time shift in the point $x_0$ in relation to the time in the center of the microparticle by

$$\Delta t = vx_0/c. \qquad (4)$$

Let us obtain the distance at which this time shift $\Delta t$ will be equal to the de Brogliee fluctuations $T_b$ (phase shift $2\pi$).

$$\Delta t = vx_0/c = T_b = h/m_0c^2; \qquad (5)$$

from where

$$x(T_b) = h/m_0v = \lambda_b. \qquad (6)$$

This way, phase shift $2\pi$ at the cost of mal-synchronization of the time, in the volume of moving microparticle from the point of view of the immobile observer, corresponds to the change of the coordinates by the length of de Broglie wave $\lambda_b$. The volume of the moving particle in the reference system of the immobile observer turns out to be phase-modulated in the direction with the spatial period $\lambda_b$, while in the microparticle's own reference system its fluctuations are synchronous and in-phase. This is the reason why the particle while interacting with the immobile object (apparatus) behaves like a wave with spatial $\lambda_b$ and different intervals of its volume interact one to another and to immobile apparatus in full accordance with its phase shifts in the reference system of the apparatus.

### 3. Phase speed of the matter waves and light speed

The second particularity of the matter waves is the phase speed of the wave that exceeds the light speed by as much as the light speed exceeds the speed of the particle moving in the space. But according to the Dirac's electron theory [9] the momentary velocity of the electron always equals the light speed, whatever its average movement speed in the space is. This conclusion from the electron theory was considered so important by Dirac, he even mentioned it in his Nobel Lecture [10]. According to Dirac's theory, the momentary velocity of the electron can have a value of only $\pm c$. At that, the electron participates simultaneously in the oscillation process with de Broglie wave frequency $\nu = m_0c^2/h$. While moving in the space with average velocity v, it fluctuates at light speed, including in the direction of movement. But what leads from it is that the concept of the phase speed is not applicable to the matter wave, because wave-particle (electron, in this case) moves not by the rectilinear but by a more complex, rather saw-tooth trajectory (Picture 1).

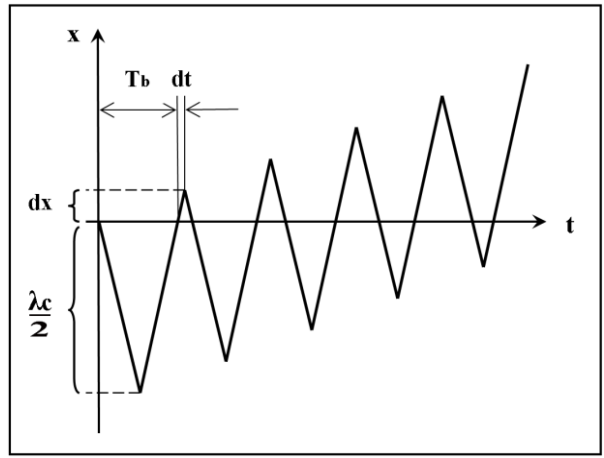

**Picture 1.** The nature of electron's movement according to Dirac's theory.

As the matter waves happen to be the form of existence not only of electron but of any other material particles, this conclusion should relate to any microparticle. So, this particularity of the matter waves is explained by the unreasonableness of implementation of the phase speed to the microparticles, while it has sense only for the rectilinear wave advance.

Dirac's theoretical conclusion about oscillatory nature of the microparticles movement at light speed is the only relation with quantum mechanics used here. Its presence allows stating that the analysis we performed earlier on the de Broglie waves is beyond the scope of classical physics. But we can return it within this scope, introducing Dirac's discovery, no matter how it was obtained, to the classical definition of the light speed, having assumed that the light speed is not only maximally possible speed of information transmission and macroscopic bodies movement but also the only possible momentary velocity of movement of the matter particles, both possessing and not possessing the rest mass.

**4. Einstein's realism and the results of verification of Bell's inequalities**

For not to get into the trap of the terminological inaccuracies let us mention that when we speak of the contradiction between the results of the experimental verification of Bell's inequalities and the requirement of the reality of the world that Einstein was insisting on, we mean not the realism in large but only its certain part that we can determine by the quote from the classic Aspec's work [6]: *«Entanglement is certainly a property beyond any space-time description by Einstein: a pair of entangled photons must be regarded as a united global object that cannot be seen as composed from separated objects with well-defined properties that are divided in time and space»*.

Let us review this pair of entangled photons. At the moment of their birth we can surely call them a united object, not considering it the violation of the world description according to Einstein. After that, reviewing them in the observer's reference system, we see these two photons flying in different directions and ending their flight in the polarizers that can be situated on different distances from the place of their birth. For example, one can be in the neighboring room, and another – in the neighboring galaxy.

Now let us see their story from the point of view of the Einstein's special theory of relativity, according to which the lifespan of the photon from the moment of its birth until it gets into the polarizer in its own reference system equals zero:

$$dt´ = dt(1-c^2/c^2)^{1/2} \equiv 0; \qquad (7)$$

where dt – time span in the observer's reference system, and dt´ - in the reference system of the photon that moves with light speed c.

That means that the moment when each photon gets in its polarizer coincides with the moment of its birth, and there is a zero time interval between them, and no change of state can happen because any physical change requires time that is more then zero. Therefore, in accordance with the special theory of relativity, in photon's reference system this pair must stay a united object during the whole flight from the source to the polarizer, however long these flights could last in our reference system.

Regarding the situation on the Aspec's example of parallel arrangements of polarizers, corresponding to the full correlation, we will obtain the following picture: if the first photon at the moment of its birth in its reference system (but in 1 microsecond in our reference system) gets into «+» channel of the polarizer, situated 300 meters apart from the source, then the second photon at the same moment of birth in its reference system (but in one year in our reference system), staying the whole entity with the first photon, with necessity will get into «+» channel of the polarizer, situated 1 light year apart from the source.

Yes, the results of experiments for Bell's inequality described by Aspect [6] in form B.C.H.S.H. complies with the inequality

$$-2\sqrt{2} \leq S(A,B) \leq 2\sqrt{2}; \qquad (8)$$

which is equivalent to the case of full correlation described above. But it does not speak about the breakdown of 'Einstein's space-time description' and not about the failure of the Einstein's realism but only about the need to make a next step in understanding of the nature of the time and space, already laid out by Einstein 100 years ago into the formulas of the special theory of relativity.

**5. Conclusion**

In the works of the authors of the theory of microparticles in its infanthood (first half of the last century), we can clearly see the deep shock of the physicists after the specifics of the microphysics were discovered. Primarily the one of de Broglie wave mechanics, which made many of them underline the differences between the microphysics and the classical physics when interpreting the results. And even claim, like Haas did, that *'the essence of the physical progress is in a gradual liberation of physics from the purely human point of views'*.

Nevertheless, when the first shock faded, and Einstein with his marvelous intuition and de Broglie himself began to insist that it is necessary to try and go back to the classical, intuitively comprehensive ways to interpret the microphysics. It is doubtless that any advancement in this direction would not only be the tribute to these great physicists of the 20th century but could also give a new impulse to the development of the theory of microparticles.

The specifics of space-time leading to the violation of bell's inequalities, requires special conditions for observation and does now manifest in out mundane life in the macroworld. But it can be considered a new wonderful instrument for the further investigation of the space-time properties to continue the left to us by Einstein [3] search of the adequate means to describe these not acknowledged yet properties of the real nature.

Planck's quantum theory that forms the basis of the theory of microparticles, tells us about a crucially quantum nature of the interaction processes between the microparticles, their creation, destruction, energy interchange. We know the worldviews (for example, [11],[12]), where it is supposed that the creation and the destruction of the microparticles is accompanied by the creation and the destruction of their individual spaces, the total of which generates what we call our usual space.

At that, we can also assume the appearance of the effects observed when studying the Bell's inequalities. If the pair of particles entangled at birth exists in the related individual spaces that are not violated from the moment of the pair's birth until the next events, then they are possibly to be regarded as a whole entity, no matter how far from each other they managed to move away in our reference system. If exterior interactions get involved in the interval between the birth of the pair of particles and them being registered in the detector, this wholeness breaks. Therefore, when there are many exterior interactions we catch the phenomenon of quantum decoherence.

Regardless the truth or untruth of the hypotheses above, the attempts to realize the manifestations of the matter movements that now seem to us unusual are still more reasonable then refusing the reality of the world under our study that leads to the statement that the moon exists only when we are looking at it.

December 2014